# Phonon calculations in cubic and tetragonal phases of $SrTiO_3$: a comparative LCAO and plane wave study

R. A. Evarestov[1], E. Blokhin[2], D. Gryaznov[2,3], E. A. Kotomin[2,3], J. Maier[2]

[1]*Department of Quantum Chemistry, St. Petersburg State University, Peterhof, Russia*

[2]*Max-Planck Institute for Solid State Research, Stuttgart, Germany*

[3]*Institute for Solid State Physics, University of Latvia, Riga, Latvia*

**Abstract**

The atomic, electronic structure and phonon frequencies have been calculated in a cubic and low-temperature tetragonal $SrTiO_3$ phases at the *ab initio* level. We demonstrate that the use of hybrid exchange-correlation PBE0 functional gives the best agreement with experimental data. The results for the standard PBE and hybrid PBE0 are compared for the two types of basis sets: a linear combination of atomic orbitals (LCAO, CRYSTAL09 computer code) and plane waves (PW, VASP 5.2 code). Relation between cubic and tetragonal antiferrodistortive (AFD) phases and the relevant AFD phase transition observed at 110 K is discussed in terms of group theory and illustrated with analysis of calculated soft mode frequences at the $\Gamma$ and $R$ points in the Brillouin zone. Based on phonon calculations, the temperature dependences of the Helmholtz free energy and heat capacity are in a good agreement with experiment.

## 1. INTRODUCTION

$ABO_3$-type perovskites continue to attract great attention due to fundamental problems of materials physics and chemistry and also numerous high tech applications [1, 2]. In particular, one of the best studied perovskites is $SrTiO_3$ (hereafter STO) which serves as a prototype for a wide class of perovskites. This incipient ferroelectric reveals the antiferrodistortive (AFD) phase transition near 110 K [3]. It was shown that the soft phonon mode at the *R*-point of the Brillouin zone (BZ) of cubic crystal is condensed below 110 K resulting in tetragonal lattice distortion with a slight unit cell stretching (see [4] and references therein) and a $TiO_6$ octahedra antiphase rotation in the nearest unit cells along the *c* axis. Additional interest to STO in 90s was fueled by a suggestion about possible ferroelectric transition around 37 K [5].

First *ab initio* calculations on STO electronic structure and phonon frequences were started in 90s (see [4, 6] and references therein). Since tetragonal structure transformation is very tiny, STO is a perfect model system for a testing of new theoretical methods. Two main techniques for phonon calculations are the *direct frozen-phonon* (DFP) [7] and the *linear-response* (LR) [8] methods, being both completely independent on any experimental data and fitted parameters. The literature analysis of the applications of these methods to a cubic STO reveals rather inconsistent results concerning the presence of the soft modes at the different BZ points (and, if present, whether the corresponding frequencies are real or imaginary).

The PW-LDA calculations performed by Sai and Vanderbilt [4] within the DFP method showed the concurrent character of AFD and FE instabilities, *i.e.* hardening of the AFD *R*-phonon and softening of the FE *Γ*-phonon with the volume increase with respect to theoretical equilibrium. The LDA, GGA-type PBE and HF-DFT hybrid HSE06 calculations performed by Wahl et al. [9] using the PW formalism within the DFP method as implemented in VASP code revealed the soft mode at the *Γ*-point being real for LDA and imaginary for the PBE and HSE06 methods. LaSota [6] and Lebedev [10] in the linear augmented plane wave (LAPW) LDA calculations within the LR method obtained imaginary soft modes at the *M*-, *R*- and *Γ*-points of the BZ. More details on the lattice dynamics of the cubic STO can be found in [11, 12, 47, 49]. Concerning the AFD phase, there is the only PW-LDA study [4] dealing with phonons. However, the AFD phase geometry obtained in this study is quite far from experiment. In fact, there is the only hybrid B3PW LCAO study [13] where the correct geometry of the STO AFD phase and its energetic preference with respect to the cubic phase was predicted. It was shown therein that the doubling of the unit cell, despite tiny tetragonal distortion, changes the indirect band gap to direct one, which is well observed in photoelectron spectra.

We performed this study using several basis sets (LCAO and PW) and DFT functionals (PBE, as well as hybrid PBE0 and B3PW) to understand their corresponding role in a description of the atomic, electronic and vibrational properties of the cubic and AFD phases of STO. This allows us to avoid uncertainties in a comparison of the results of above-mentioned different calculations when even the small differences in the precision (*e.g.* choice of cut-off energies or *k*-point mesh) could be important. Furthermore, this work was also stimulated by the absence of the *ab initio* thermodynamic comparison of different STO phases in literature.

This paper is organized in a following way. In Section 2 the group-theory analysis of the cubic and AFD phase and relevant phase transformation is performed. The computational details are discussed in Section 3. The main results for atomic and electronic structure and phonon frequencies obtained for two STO phases are summarized in Section 4 along with the calculations of the heat capacities and the Helmholtz free energies. Lastly, the conclusions are summarized in Section 5.

## 2. PHONON SYMMETRY IN CUBIC AND TETRAGONAL PHASES

The STO cubic phase (space group $Pm\bar{3}m$ $O_h^1$) represents an example of the ideal perovskite $ABO_3$ structure. We set the octahedrally coordinated B cation at the Wyckoff position 1a (0, 0, 0), the A cation at 1b (1/2, 1/2, 1/2), and the anion X at 3d (1/2, 0, 0) (Table 1 and Fig. 1a ). To analyze the symmetry of phonon states, the method of induced representations (hereafter reps) of space groups can be used [14, 15]. The total dimension *n* of the induced rep (called the mechanical rep) equals *3N* (*N* is the number of atoms in the primitive cell and equals 5 for ideal perovskite).

Table 1 shows the phonon symmetry in the STO crystal for $\Gamma$ (0, 0, 0) and $R$ (1/2, 1/2, 1/2) symmetry points of the BZ for a simple cubic lattice. The space group $O_h^1$ irreducible reps (irreps) are labeled according to Ref. [16]. The space group irreps are induced from those site symmetry group irreps, which correspond to transformations of the atomic displacements (x, y, z): $t_{1u}$ of the site symmetry group $O_h$ (Ti and Sr atoms); $a_{2u}$, $e_u$ of the site symmetry group $D_{4h}$ (O atom).

One can obtain four $t_{1u}$ modes and one $t_{2u}$ mode at the $\Gamma$ point of the BZ; one $t_{1u}$ mode is acoustic. Three phonon modes of the $t_{1u}$ symmetry are infrared (IR) active and one mode of the $t_{2u}$ symmetry is neither infrared nor Raman active (so called "silent mode"). The latter mode is connected with the displacements of O atoms only. Three modes $1^+$, $3^+$ and $4^+$ at the $R$-point of the BZ (with the degeneracy 1, 2 and 3, respectively) are displacement modes of O atom only. The 3-fold degenerated $R_{4-}$ and $R_{5+}$ modes are Ti and Sr-O modes, respectively.

The STO structural phase transition at 110 K reduces the symmetry from a cubic to tetragonal (space group *I4/mcm*-$D_{4h}^{18}$). The primitive unit cell of the body-centered tetragonal lattice consists of 10 atoms (the cubic unit cell is doubled). Fig. 1b shows the crystallographic (quadruple) unit cell of body-centered tetragonal lattice and the occupations of Wyckoff positions by atoms. It is seen that O atoms are separated in two nonequivalent orbits: 4a (two atoms) and 8h (four atoms). The quadruple unit cell parameters in the undistorted pseudo-cubic structure are $a = b = a_0\sqrt{2}$ , $c = 2a_0$, where $a_0$ is the lattice constant of the cubic phase. The structural parameter $u = 0.25$ defines the oxygen 8h position in the undistorted pseudo-cubic structure. Thus the experimental cubic lattice parameter $a_0 = 3.905$ Å (at room temperature) gives the undistorted pseudo-cubic structure with $a = b = 5.522$ Å, $c = 2a_0 = 7.810$ Å which are close to the experimental $a = 5.507$ Å, $c = 7.796$ Å [43] for a real distorted tetragonal structure. The experimental value $u = 0.241$ [43] is close to $u = 0.25$ for the undistorted pseudo-cubic structure. Thereby the cubic-to-tetragonal phase transition can be considered as the tetragonal supercell generation with the transformation matrix $\begin{bmatrix} 110 \\ -110 \\ 002 \end{bmatrix}$ and the further small structural distortion.

The $BO_6$ octahedra distortions are considered often in terms of tilting (see [18] and references therein). By this one means the tilting around one or more of the $BO_6$ symmetry axes allowing greater flexibility in the coordination of the A cation while leaving the environment of the B cation essentially unchanged. The tilt of one octahedron around one of these axes determines (via the corner connections) the tilts of all the octahedra in the plane perpendicular to this axis. The successive octahedra along the axis can be tilted in either the same or the opposite sense (in-phase and out-of-phase tilts). The group-theoretical analysis of octahedral tilting is described in detail in Ref.[18]. It was shown that the irrep associated with the out-of -phase tilts is $R_{4+}$

The symmetry of phonons in the AFD STO phase at the $\Gamma$ point of the BZ is shown at Table 1 along with the splitting of the phonon frequencies due to lowering of the symmetry. The symmetry of acoustic phonons is ($a_{2u} + e_u$), 8 modes are IR active (3 $a_{2u}$ + 5 $e_u$), 7 modes are Raman active ($a_{1g}$, $b_{1g}$, 2 $b_{2g}$, 3 $e_g$). The silent modes have the symmetry $a_{1u}$ and $b_{1u}$ in IR spectra and 2 $a_{2g}$ in Raman spectra.

As it is seen from Table 1, $a_{1g}$ and $b_{1g}$ Raman active phonons arise due to the displacements of the O atoms only. The Raman active modes with the symmetry $b_{2g}$ and $e_g$ are Sr-O vibrational modes. The vibrations connected with Ti atom displacements are active only in IR spectra ($a_{2u}$, 5 $e_u$ phonons). The information about the connection of the active vibrational modes with the displacements of concrete atoms makes the interpretation of the experimental IR and Raman spectra easier (see Section 4).

The calculations of phonon frequencies in a cubic high temperature phase (presented below) clearly demonstrates that the soft phonon mode symmetry is $R_{4+}$ and in this way confirms the results of the group-theoretical analysis. In the next sections we discuss the computational details of the present phonon calculations and compare them with those found in the experimental and other theoretical studies.

## 3. CALCULATION DETAILS

In this study, the DFP method was chosen for phonon calculations as it can, unlike the LR method, be used in conjunction with any external atomistic simulation code (ability to compute forces is only required). To obtain the phonon frequencies and thermodynamic functions within the DFP method, three steps are required [19]: (1) structure optimization, (2) constructing of supercell based on optimized structure and displacing atoms inside it, (3) calculation of induced forces and the corresponding force matrix diagonalization.

At the first step, the total energy minimization is performed in order to find the equilibrium atomic structure of the crystal. For this purpose we used CRYSTAL09 [20] and VASP 5.2 [21] computer codes for checking both the LCAO and PW schemes (see *ab initio* calculation details below). One should notice that not every wave vector *k* commensurates with any supercell. For a cubic STO we used the 2x2x2 supercell of 40 atoms since it is small enough to be calculated in a reasonable time but large enough to commensurate with all the symmetry *k*-points of the BZ (*Γ*, *R*, *X*, *M*) for a simple cubic lattice. For the convenience of comparison, the same 40-atoms supercell was used also for the AFD STO phase.

The calculation of forces induced by displaced atoms is the third step in phonon calculation. In practice, the weak point of the DFP method is the fact that the magnitude of atomic displacements are arbitrary in the force calculation and the computer codes implementing this method could produce inconsistent results depending on a choice of this magnitude. Generally, the atomic displacements should not be too large, to guarantee the linear relation between forces and displacements (as the harmonic approximation is used in DFP method). However, these displacements also should not be too small to avoid an effect of a numerical noise in the total energy second derivative calculation and therefore redundant calculation accuracy. Our experience shows that for STO crystal the default value of displacement magnitude of 0.003 Å in CRYSTAL09 code is reasonable. However, the same magnitude in VASP 5.2 code seems to be insufficient as we obtained anomaly large acoustic phonon frequencies with this magnitude. The value of displacement magnitude in PW-based phonon calculations was fixed as 0.02 Å.

The forces thus obtained are collected in a dynamical matrix where the eigenvalues give squared phonon frequencies and eigenvectors are equal to phonon modes. Unlike VASP 5.2 code, CRYSTAL09 code allows to solve the dynamical matrix for all the symmetry *k*-points commensurating with the supercell chosen. VASP 5.2 code permits to calculate the frequencies only at the $\Gamma$ point of the BZ independently on the supercell chosen (whereas the classification of the calculated frequencies over other *k*-points is additionally required).

In order to obtain the temperature dependence of the Helmholtz free energy and heat capacity, the integration over the phonon density of states is finally performed. The corresponding expressions for these thermodynamic functions include phonon frequencies through hyperbolic functions [22]. It means that the lower the frequency, the greater its contribution. That is why one has to treat the low soft phonon modes very accurately, in order to obtain the reasonable thermodynamic functions. One should also mind that the DFP calculations neglect anharmonic effects and this is why the obtained soft mode frequences and temperature dependencies of the Helmholtz free energy and heat capacity are correct only qualitatively.

In the *ab initio* calculations of phonon properties using CRYSTAL09 (the LCAO basis set) and VASP 5.2 (the PW basis set) we have chosen the DFT GGA-type PBE functional [23] and two hybrid (PBE0 [24] and B3PW [25]) exchange-correlation functionals (currently B3PW is not implemented in VASP 5.2). All these functionals have been successfully applied earlier for the calculations of bulk and surface perovskite properties [9, 13, 26].

In all the LCAO calculations the small-core relativistic effective core potentials (ECP) for Ti and Sr atoms [27, 28] were used, while the all-electron triple-zeta (TZ) quality basis set (BS) for O-atom has been taken from Ref. [29].

It is well known that in the LCAO calculations of crystals the BS of a free atom has to be modified as the diffuse outmost wave functions cause numerical problems because of the large overlap with the core functions of the neighboring atoms in a dense-packed crystal [30]. To optimize the BS in present study, we used the Powell's conjugate-directions minimization method [31] without calculations on the total energy derivatives. It is known as one of the most efficient direct minimization methods. Being interfaced with CRYSTAL09 code, our computer program OPTBAS [32] has been applied for the basis set optimization. The Gaussian-type orbital (GTO) exponents less than 0.1 $bohr^{-2}$ were excluded and the bound-constrained optimization has been performed for the remaining exponents with 0.1 $bohr^{-2}$ lower bound. The GTO exponents in non-contracted basis functions [27, 28] were optimized in the PBE0 calculations for the cubic bulk STO and compared with non-optimized ones in Table 2. One should note that the first attempt to optimize the LCAO basis set for STO has been performed only recently in B3PW calculations [26].

The following precision settings were applied in both CRYSTAL09 and VASP 5.2 codes, unless otherwise stated. The Monkhorst–Pack [33] 8×8×8 *k*-point mesh in the BZ was used. The tolerance on the energy convergence on the self-consistent field (SCF) cycles was set to $10^{-10}$ a.u. and an extra-large pruned DFT integration grid was adopted. In CRYSTAL09 code the truncation criteria for bielectronic integrals (Coulomb and HF exchange series) were heightened (values 8, 8, 8, 8, 16 [20]). Additionally the DFT density and grid weight tolerances were heightened (values 8 and 16 [20]). In PW framework we used the projector augmented wave (PAW) method [34] and scalar-relativistic ECPs substituting for 28 core electrons on Sr atom, 10 core electrons on Ti atoms and 2 core electrons on O atoms. The plane wave cut-off energy was fixed at 600 eV for both the geometry optimization and the phonon frequency calculations. The electron occupancies in VASP 5.2 were determined with the Gaussian method using a smearing parameter of 0.1 eV.

## 4. RESULTS AND DISCUSSION

### A. Cubic phase

The basic bulk properties of STO in a cubic phase, calculated using both the LCAO and PW basis sets and PBE, PBE0 and B3PW functionals, are presented in Table 3 along with the experimental data. As one can see, the PBE functional in both the LCAO and PW calculations considerably underestimates the band gap and overestimates the lattice constant. (This is well-known trend in GGA calculations, in general, and has been observed also for STO, in particular in Ref. [26].) Moreover, the band gap for the same PBE functional within the PW basis set becomes even more underestimated than in the LCAO calculations, indicating an importance of basis set flexibility. This is confirmed by a comparison of the results for the PBE0 functional with optimized and non-optimized basis sets: the optimization improves agreement with the experiment not only for the band gap (electronic properties), but also for crystal atomization energy and bulk modulus (thermochemical and mechanical properties). One should note that the B3PW hybrid functional also gives very good results for the most properties.

The calculated TO and LO phonon frequencies for the cubic phase at the $\Gamma$ and $R$ points of the BZ are summarized in Tables 4 and 5. To calculate phonons in these points, the primitive unit cell should be doubled with the following transformation matrix $\begin{pmatrix} 110 \\ 101 \\ 011 \end{pmatrix}$ of the lattice translation vectors [30].

The upper part of the Table 4 present results for the $\Gamma$-point calculations. All the methods used predict the $t_{1u}$ TO1 mode to be soft, in agreement with the experiment [2, 3]. Its frequency is either imaginary or very low, depending on the particular functional. Moreover in all the PW calculations it is imaginary. (It is important to note that our calculations are performed in the harmonic approximation and thus unable to obtain correctly the soft mode frequencies.) Numbers in brackets for the PBE0 functional show relative errors in frequency calculations, $|\nu_{exp}-\nu_{th}| / \nu_{exp} \cdot 100\%$. As a rule, the largest errors are observed for the low frequencies but agreement is much better for the large frequencies. Similarly, basis set optimization in the LCAO approach greatly improves the results and reduces the errors by a factor of 2-5 as compared with the case of non-optimized one.

The lower part of the Table 4 presents the phonon frequencies at $R$ point of the BZ. We found no other relevant calculations in the literature. In agreement with the group theoretical analysis (Section 2) and inelastic neutron scattering experiments [40], we found instability for the $R_{4+}$ soft mode. All six calculated frequencies within the PBE0 functional are in good agreement with the experiment (the largest error for non-soft modes is only 5%). Also the PW calculations with the PBE0 functional suggest high quality results, the relative errors do not exceed 7%. Our results for the PW PBE0 calculations are very similar to those recently published for the HSE06 hybrid functional with screened Coulomb interactions [9].

The dependence of the soft mode frequencies at the $\Gamma$ and $R$ points are plotted in Fig.2 as a function of the lattice constant (which mimics thermal expansion of the lattice). As one can see, the soft mode at the $R$ point increases very slowly, whereas that at the $\Gamma$ point decreases considerably, supporting the idea [4] about their concurrent character. The recent simulations using the ABINIT code with GGA-PW91 functional [47] have shown the same trend but with an imaginary frequency at the *R* point at all lattice parameters calculated.

The agreement with the experiment for three LO frequences (Table 5) is also very good, as well as magnitude of the LO-TO splitting. The earlier PW calculations within LDA approximation [4] gave worse results than our hybrid-functional calculations.

## B. Tetragonal AFD phase

The basic structural properties for the tetragonal AFD phase are presented in Table 6 for the LCAO with three different functionals and PW with PBE functional (we did not perform PW calculations with hybrid functionals as they are extremely time-demanding). Along with the two lattice constants *a* (= *b)* and *c*, O-ion position parameter *u* and the relevant $TiO_6$ octahedra rotation angle are compared. Firstly, the LCAO calculations with PBE0 functional and optimized basis give the best agreement with the experiment. Secondly, the PW calculations strongly overestimate the lattice constants and tetragonal lattice distortion (see also [4]).

We analyzed the two phase energies in details in Table 7. Two hybrid functionals, B3PW and PBE0, give moderate total electronic energy gain ΔE for the tetragonal AFD phase with respect to the cubic one, whereas for the PBE this gain is only 0.6 meV. It is necessary also to take also into account the zero-point vibration energies $\Delta E_{ZP}$ (second row) which results in a small AFD final energy gain (not exceeding 20 meV) for all the functionals used within LCAO scheme. Calculating the zero-point energies, we compared values for the cubic and tetragonal AFD supercells with the *same* number of atoms. The temperature dependence of the Helmholtz free energies based on frequences will be discussed below.

Lastly, the phonon frequencies for the AFD phase are presented in Table 8 where the theoretical predictions for Raman, infrared and silent modes are compared with the experimental and theoretical PW-LDA [4] data. The main conclusions are the following. (i) The PBE0 functional gives the best agreement with the experimental data (when available). (ii) The splitting of three cubic $t_{1u}$ modes (denoted by curved brackets in Table 8) are predicted to be small, 2, 4 and 11 $cm^{-1}$. (iii) Several soft modes are found indicating a possible instability of tetragonal AFD phase at the temperatures close to 0 K as discussed in Ref. [4]. As before, an agreement with the experiment for the high frequencies is better than for the low frequencies.

### C. Heat capacity and Helmholtz free energy

The temperature dependence of heat capacity $C_v$ at fixed volume and pressure has been calculated for two STO phases using PBE, PBE0 and B3PW functionals (only the second is shown at Fig.3). In all three cases the agreement with experiment [48] is excellent, and $C_v$ curves for the cubic and AFD phases lie very close in a wide temperature range. To the best of our knowledge, the only heat capacity calculations [50] were performed so far for STO cubic phase using the all-electron LAPW method (WIEN-2k code) with the results being very close to our calculations.

Using CRYSTAL09 code, we calculated also the Helmholtz free energy for PBE0 functional as the function of the temperature (Fig.4). The energy curve for the tetragonal AFD phase lies slightly below the curve for the cubic phase in a whole temperature range. This contradicts to the experimental fact that these two curves cross around the phase transition temperature (110 K). We assume that this contradiction results from inability of the mean-field quantum electronic structure methods and harmonic approximation used here to reproduce tiny details of phase transitions.

## 5. SUMMARY

A comparison of the phonon frequences and atomic structure of the cubic and tetragonal AFD STO phases calculated within the LCAO approach using optimized and non-optimized basis sets clearly demonstrates that basis set optimization gives much better results.

The use of the PBE0 hybrid functional with both types of basis sets, LCAO and PW, gives better phonon frequences compared to GGA functionals. Moreover, the LCAO hybrid DFT calculations in CRYSTAL09 are much faster compared with extremely time consuming hybrid DFT calculations in VASP 5.2, when obtaining even the bulk properties for tetragonal AFD phase becomes very time-consuming.

The splitting of the phonon frequences $t_{1u} \rightarrow a_{2u} + e_u$ due to AFD phase transition is predicted to be rather small, 2-11 $cm^{-1}$. Our calculations confirm co-existence of the soft phonon modes for the cubic phase at the $\Gamma$ and $R$ points of the BZ which was intensively discussed earlier [4, 6, 10, 28]. Lastly, the experimental temperature dependence of the STO heat capacity is successfully reproduced.

Based on this experience for defect-free STO, we plan to use the LCAO approach combined with hybrid functionals for further thermodynamical study of defective perovskites under finite temperatures, which is important for prediction materials properties and device performance (e.g. sensors and solid oxide fuel cells) under realistic operational conditions.

**Acknowledgements**

Authors are greatly indebted to M. Cardona, V. Trepakov, E. Heifets, R. Merkle, J. Gavartin for many stimulating discussions. This study was partly supported by ERANET MATERA project.

**Figure captions**

Fig.1. Cubic (a) and AFD (b) STO

Fig.2. Γ and R soft phonon modes vs. temperature dependence calculated via PBE0 (opt. BS) / LCAO scheme

Fig.3. Heat capacity calculated via PBE0 (opt. BS) / LCAO scheme

Fig.4. Helmholtz free energy calculated via PBE0 (opt. BS) / LCAO scheme

Table 1.
Wyckoff positions and phonon symmetry in cubic and tetragonal AFD $SrTiO_3$

| $Pm\bar{3}m$ $O_h^1$ (SG 221) | | | $I4/mcm$ $D_{4h}^{18}$ (SG 140) | |
|---|---|---|---|---|
| | Γ | R | | Γ |
| **Ti** | | | **Ti** | |
| **1a (0, 0, 0)** | | | **2c (0 0 0)** | |
| $O_h$ $t_{1u}$ (x, y, z) | $4^-$ ($t_{1u}$) | $4^-$ | $C_{4h}$ $a_u$ (z) | $a_{1u}$ $a_{2u}$ |
| | | | $e_u'$ (x, y), $e_u''$ (x, y) | $2e_u$ |
| | | | | $t_{1u} \rightarrow a_{2u}\ e_u$ |
| **Sr** | | | **Sr** | |
| **1b (0.5, 0.5, 0.5)** | | | **2b (0 0.5 0.25)** | |
| $O_h$ $t_{1u}$ (x, y, z) | $4^-$ ($t_{1u}$) | $5^+$ | $D_{2d}$ $b_2$ (z) | $a_{2u}$ $b_{2g}$ |
| | | | | $R_5^+ \rightarrow b_{2g}\ e_g$ |
| | | | $e$ (x, y) | $e_g$ $e_u$ |
| | | | | $t_{1u} \rightarrow a_{2u}\ e_u$ |
| **O** | | | **O** | |
| **3d (0.5 0 0)** | | | **2a (0 0 0.25)** | |
| $D_{4h}$ $a_{2u}$ (z) | $4^-$ ($t_{1u}$) | $1^+3^+$ | $D_4$ $a_2$ (z) | $a_{2g}$ $a_{2u}$ |
| | | | $e$ (x, y) | $e_g$ $e_u$ |
| $e_u$ (x, y) | $4^-$ ($t_{1u}$) | $4^+5^+$ | | $R_1^+ \rightarrow a_{2g}$ |
| | $5^-$ ($t_{2u}$) | | | $R_3^+ \rightarrow e_g$ |
| | | | | $t_{1u} \rightarrow a_{2u}\ e_u$ |
| | | | **O** | |
| | | | **4h (−*u*+0.5, *u*, 0)** | |
| | | | $C_{2v}$ $a_1$ (z) | $a_{1g}$ $b_{2g}$ $e_u$ |
| | | | $b_1$ (x) | $b_{1g}$ $a_{2g}$ $e_u$ |
| | | | $b_2$ (y) | $a_{2u}$ $b_{1u}$ $e_g$ |
| | | | | $R_4^+ \rightarrow b_{1g}\ e_g$ |
| | | | | $R_5^+ \rightarrow b_{2g}\ e_g$ |
| | | | | $t_{1u} \rightarrow a_{2u}\ e_u$ |
| | | | | $t_{2u} \rightarrow b_{1u}\ e_u$ |

Table 2.
Optimization of GTO outer exponents ($bohr^{-2}$)

| Type | Non-opt. | Opt. |
|---|---|---|
| Sr (small core) $4s^2 4p^6 5s^2$ | | |
| 6s | 0.3344 | 0.3452 |
| 5p | 0.7418 | 0.7296 |
| 6p | 0.2801 | 0.2757 |
| Ti (small core) $3s^2 4p^6 3d^2 4s^2$ | | |
| 5s | 0.5128 | 0.6149 |
| 5p | 0.3982 | 0.3426 |
| 4d | 0.8002 | 0.7399 |
| 5d | 0.2620 | 0.2936 |
| O (all electron) $1s^2 2s^2 2p^4$ | | |
| 3sp | 0.9057 | 0.8785 |
| 4sp | 0.2556 | 0.2287 |
| 3d | 1.2920 | 0.1480 |

Table 3.
Cubic STO basic properties

| | LCAO | | | | PW | | Expt. |
|---|---|---|---|---|---|---|---|
| | PBE | PBE0 | | B3PW | PBE | PBE0 | |
| | | non-opt.BS | opt.BS | | | | |
| Lattice constant $a_0$, Å | 3.958 | 3.890 | 3.913 | 3.924 | 3.944 | 3.902 | 3.905 [35] |
| Band gap, eV | 1.8 | 4.0 | 3.9 | 3.4 | 1.1 | 3.5 | 3.3 [36] |
| Atomization energy, eV | 31.6 | 28.5 | 29.3 | 29.1 | 32.8 | 32.2 | 31.7 [37] |
| Bulk modulus, GPa | 171 | 220 | 195 | 190 | 169 | 193 | 179 [38] |

Table 4.
TO phonon frequencies ($cm^{-1}$) in cubic STO phase
*(numbers in brackets are relative errors in percents with respect to expt. [40])*

| | | LCAO | | | | PW this work | | PW [9] | | Expt. (297 K) [39][a], [40][b], [41][b] |
|---|---|---|---|---|---|---|---|---|---|---|
| | | PBE | PBE0 non-opt.BS | PBE0 opt.BS | B3PW | PBE | PBE0 | PBE | HSE06 | |
| Γ | $t_{1u}$ (TO1) | ***71i*** | 63 *(31)* | 72 *(21)* | 17 | ***133i*** | ***100i*** | ***115i*** | ***74i*** | 42 [39], 91 [40] |
| | $t_{1u}$ (TO2) | 166 | 203 *(19)* | 180 *(6)* | 175 | 146 | 161 *(5)* | 147 | 162 | 175 [39], 170 [40] |
| | $t_{2u}$ | 247 | 302 *(14)* | 271 *(2)* | 267 | 226 | 252 *(5)* | 234 | 250 | 265 [40] |
| | $t_{1u}$ (TO3) | 522 | 594 *(9)* | 547 *(0)* | 540 | 508 | 536 *(2)* | 512 | 533 | 545 [39], 547 [40] |
| R | $R_{4+}$ | ***16i*** | 92 *(77)* | 70 *(35)* | 55 | ***86i*** | ***54i*** | – | – | 52 [40] |
| | $R_{5+}$ | 144 | 177 *(22)* | 153 *(6)* | 149 | 128 | 138 *(5)* | | | 145 [40] |
| | $R_{4-}$ | 432 | 481 *(8)* | 460 *(3)* | 454 | 413 | 442 *(1)* | | | 446 [40] |
| | $R_{5+}$ | 437 | 493 *(10)* | 465 *(3)* | 461 | 419 | 449 *(0)* | | | 450 [40] |
| | $R_{3+}$ | 440 | 533 *(12)* | 478 *(1)* | 466 | 433 | 475 *(0)* | | | 474 [40] |
| | $R_{1+}$ | 804 | 906 | 861 | 848 | 798 | 857 | | | ~800 [41] |

[a] IR experiment
[b] Inelastic Neutron Scattering (INS) experiments

Table 5.
LO phonon frequencies (cm$^{-1}$) in cubic STO phase
*(numbers in brackets are relative errors in percents with respect to expt. [39])*

| | LCAO | | | | PW LDA [42] | Expt. [39] |
|---|---|---|---|---|---|---|
| | PBE | PBE0 | | B3PW | | |
| | | non-opt.BS | opt.BS | | | |
| $t_{1u}$ (LO1) | 165 | 203 *(19)* | 180 *(5)* | 174 | 158 | 171 |
| $t_{1u}$ (LO2) | 458 | 530 *(12)* | 480 *(1)* | 477 | 454 | 474 |
| $t_{1u}$ (LO3) | 833 | 810 *(2)* | 809 *(2)* | 809 | 829 | 795 |

Table 6.
AFD STO structural properties

| | | LCAO | | | PW | Expt. |
|---|---|---|---|---|---|---|
| | | PBE | PBE0 opt.BS | B3PW | PBE | |
| Lattice constants, Å | *a* | 5.594 | 5.532 | 5.545 | 5.566 | 5.507 (*50K* [43]) |
| | *c* | 7.922 | 7.831 | 7.854 | 7.908 | 7.796 (*50K* [43]) |
| Cubic-tetragonal distortion $c/(\sqrt{2}a)$ | | 1.0014 | 1.0011 | 1.0014 | 1.0046 | 1.0010 (*50K* [43])<br>1.0006 (*65−110K* [44]) |
| O atom position, *u*, frac. | | 0.245 | 0.246 | 0.245 | 0.228 | 0.240 (*4K* [3])<br>0.241 (*50K* [43])<br>0.244 (*77K* [3]) |
| $TiO_6$ rotation angle $arctg(1-4u)$, ° | | 1.1 | 0.9 | 1.1 | 4.9 | 2.1 (*4K* [3])<br>2.0 (*50K* [43])<br>1.4 (*77K* [3]) |

Table 7.
Total electronic energy difference ΔE and zero-point energy difference $\Delta E_{ZP}$ (meV per unit cell) of cubic and tetragonal AFD STO phases with respect to the cubic phase, calculated via LCAO method

| | PBE | PBE0 opt.BS | B3PW |
|---|---|---|---|
| ΔE | −0.6 | −2.9 | −1.8 |
| $\Delta E_{ZP}$ | −12.4 | −16.7 | −14.5 |
| Sum. | −12.8 | −19.6 | −16.3 |

Table 8.
Phonon frequencies (cm$^{-1}$) in AFD STO phase
*(numbers in brackets are relative errors in percents with respect to expt. [45] and [46])*

| | | LCAO, CRENBL BS | | | | | PW | PW | Expt. |
|---|---|---|---|---|---|---|---|---|---|
| | | PBE | PBE0 | | B3PW | | PBE | LDA | (15 K) |
| | | | non-opt.BS | opt.BS | | | | [4] | |
| Raman | $a_{1g}$ | 29 | 78 | 63 | 61 | | 98 | − | 48 [45] |
| | | | *(63)* | *(31)* | | | | | |
| | $e_g$ | 48 | 99 | 79 | 76 | | ***17i*** | | 15 [45] |
| | | | *(560)* | *(427)* | | | | | |
| | $e_g$ | 137 | 168 | 146 | 144 | | 183 | | 143 [45] |
| | | | *(17)* | *(2)* | | | | | |
| | $b_{2g}$ | 152 | 181 | 158 | 157 | | 140 | | 235 [45] |
| | | | *(23)* | *(33)* | | | | | |
| | $b_{2g}$ | 441 | 534 | 466 | 462 | | 421 | | |
| | $e_g$ | 444 | 537 | 468 | 465 | | 425 | | 460 [45] |
| | | | *(17)* | *(2)* | | | | | |
| | $b_{1g}$ | 438 | 501 | 479 | 469 | | 437 | | |
| | $a_{2g}$ | 440 | 502 | 480 | 470 | | 434 | | |
| | $a_{2g}$ | 806 | 908 | 862 | 850 | | 793 | | |
| Silent | $b_{1u}$ | 252 | 308 | 275 | 271 | | 245 | | |
| | $a_{1u}$ | 430 | 478 | 458 | 452 | | 410 | | |
| Infrared | $a_{2u}$ | 2 | 101 | 68 | 28 | } | ***4i*** | ***90i*** | |
| | $e_u$ | 1 | 103 | 72 | 45 | } | ***28i*** | ***96i*** | |
| | $e_u$ | 163 | 199 | 177 | 174 | } | 183 | | |
| | $a_{2u}$ | 180 | 211 | 189 | 187 | } | 158 | 157 | 185 [46] |
| | | | *(14)* | *(2)* | | | | | |
| | $e_u$ | 252 | 306 | 273 | 270 | | 239 | 240 | |
| | $e_u$ | 433 | 481 | 460 | 455 | | 411 | 419 | 450 [46] |
| | | | *(7)* | *(2)* | | | | | |
| | $e_u$ | 523 | 597 | 549 | 542 | } | 504 | 515 | |
| | $a_{2u}$ | 526 | 599 | 551 | 544 | } | 510 | | |

Fig.1.

**SG 221: Pm3m**, $T_{exp.}$>105K
$a_0$ = 3.905 Å

**SG 140: I4/mcm**, $T_{exp.}$<105K
$\sqrt{2}a_0 \approx a$ = 5.507 Å, $2a_0 \approx c$ = 7.796 Å
oxygen distortion in 8h, $u$=0.241
vertical cell stretching, $c/(\sqrt{2}a)$=1.001

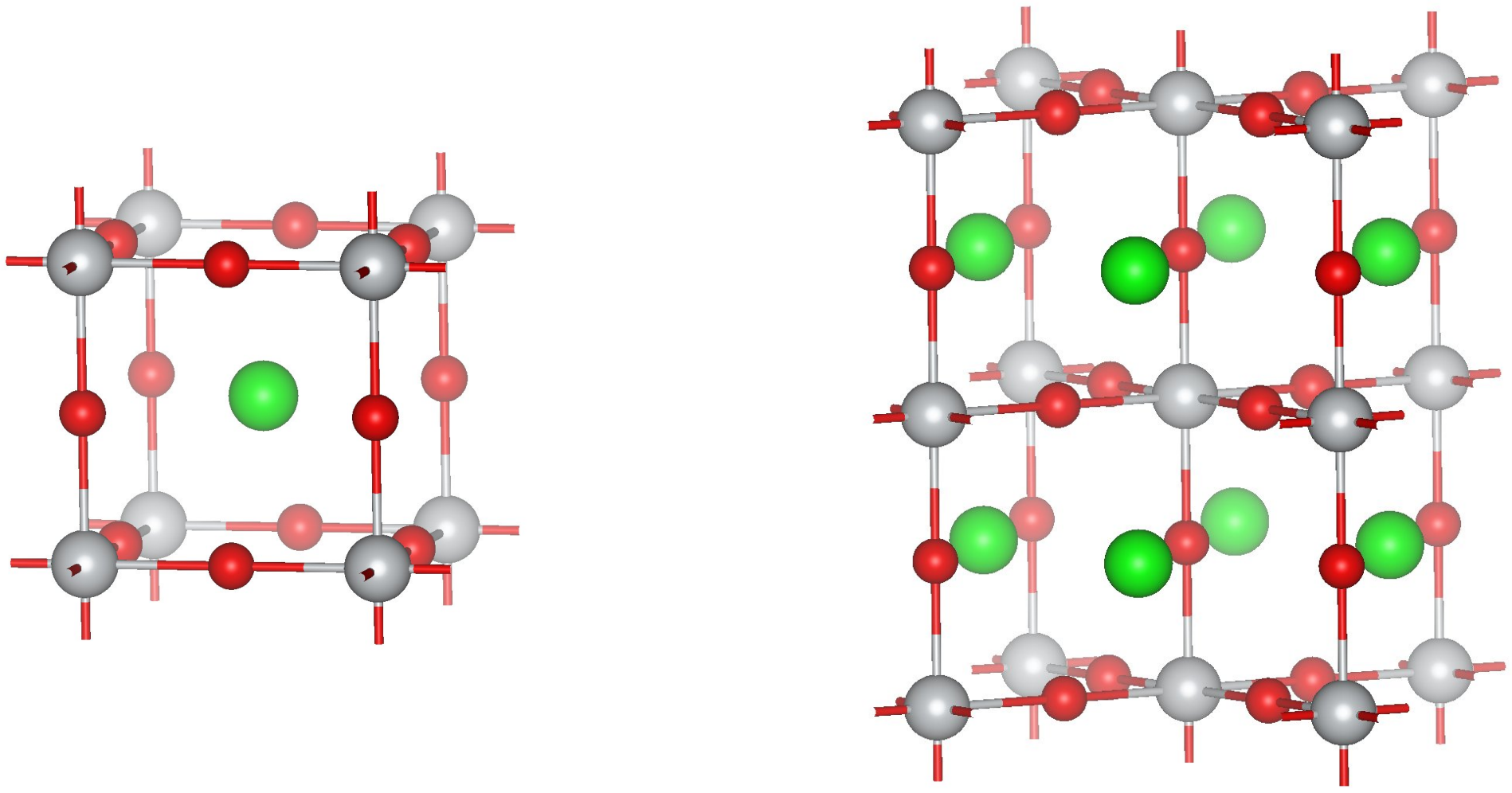

(a)

| | | | | |
|---|---|---|---|---|
| **Sr** | 0.5 | 0.5 | 0.5 | (1b) |
| **Ti** | 0.0 | 0.0 | 0.0 | (1a) |
| **O** | 0.5 | 0.0 | 0.0 | (3d) |

(b)

| | | | | |
|---|---|---|---|---|
| **Sr** | 0.00 | 0.50 | 0.25 | (4b) |
| **Ti** | 0.00 | 0.00 | 0.00 | (4c) |
| **O** | 0.00 | 0.00 | 0.25 | (4a) |
| **O** | $u$ | $u$+0.50 | 0.00 | (8h) |

Fig.2.

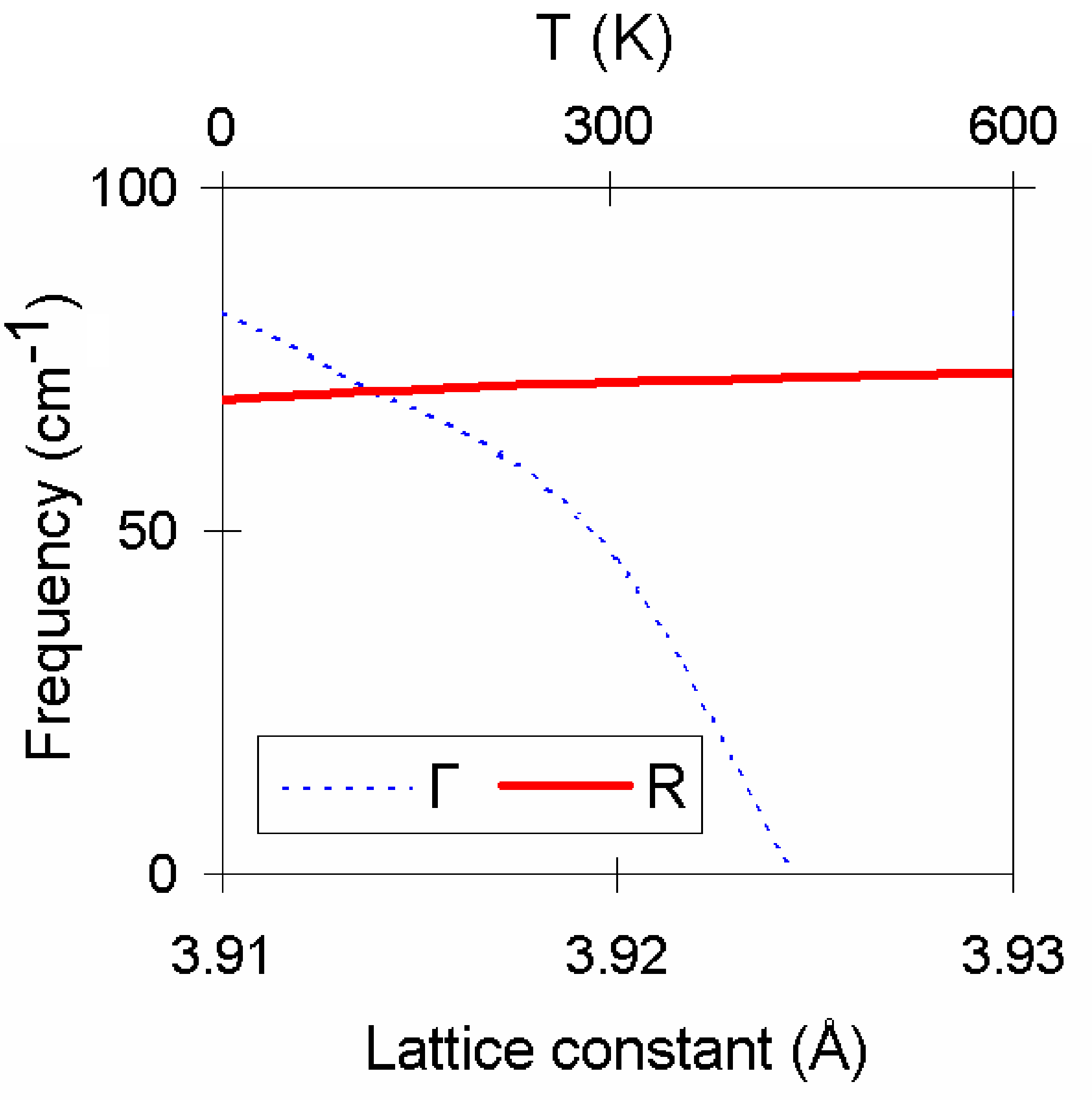

Fig.3.

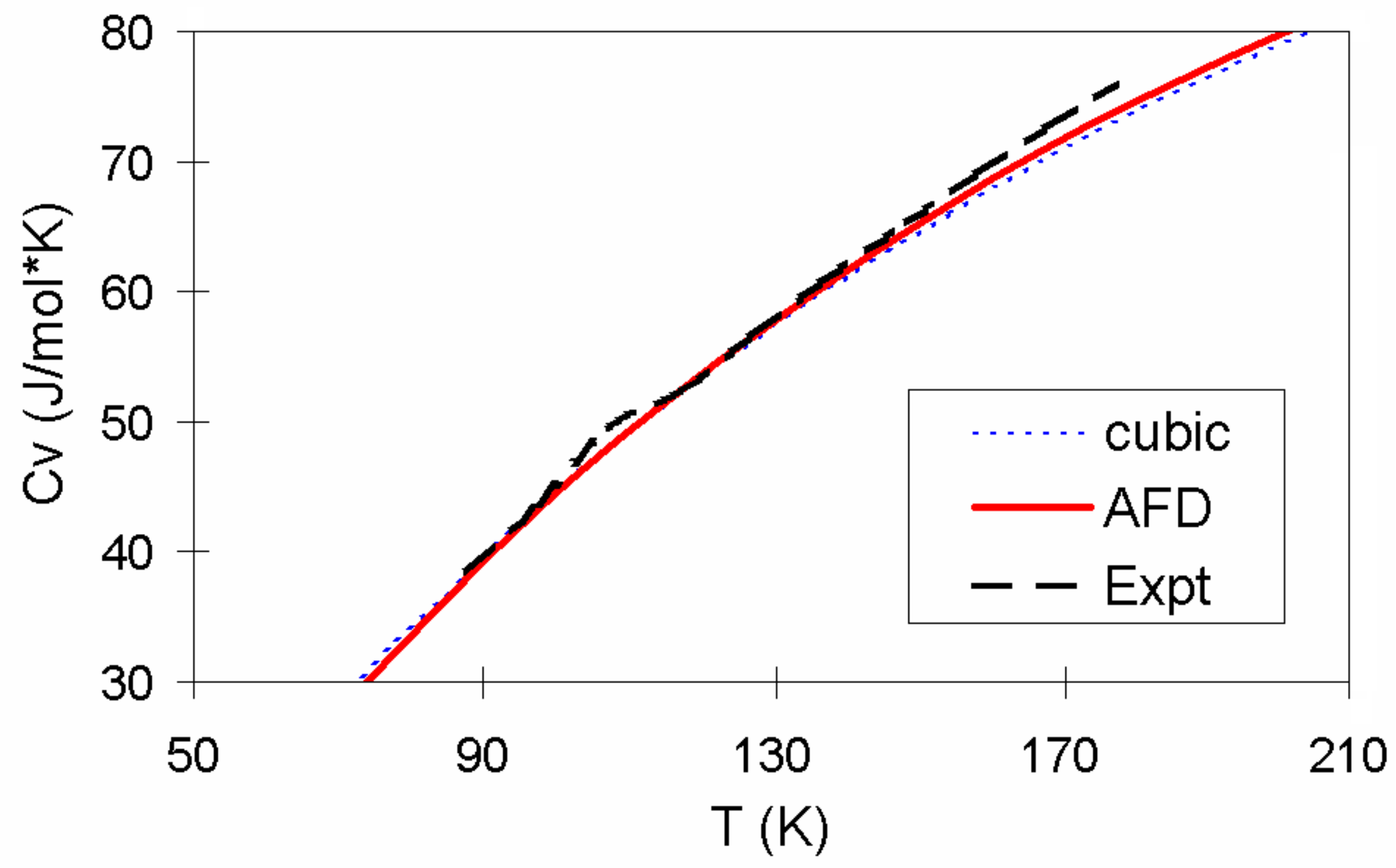

Fig.4.

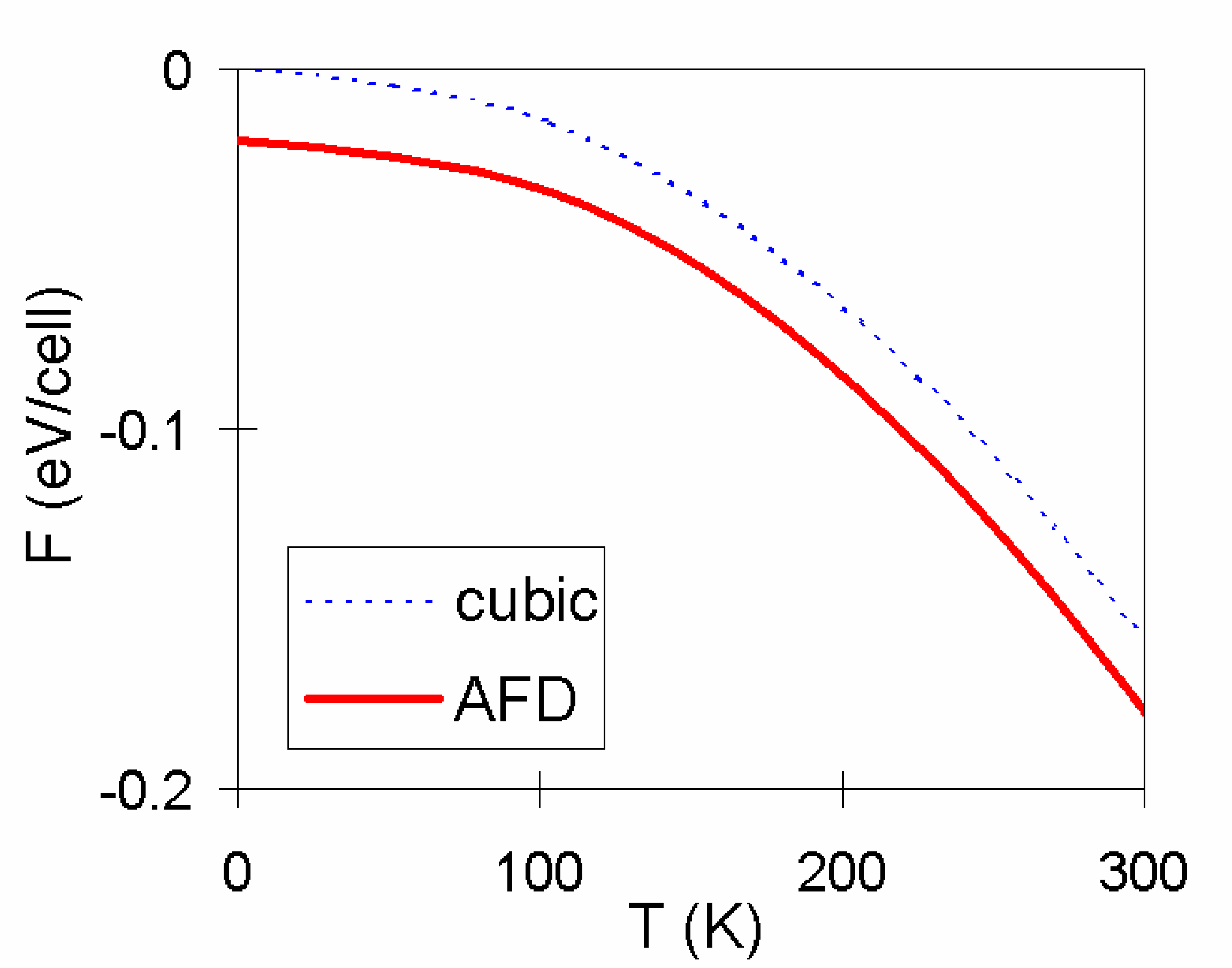